\documentclass[]{aastex62}

\received{May 28, 2019}
\revised{ZZZ YY, 20XX}
\accepted{\today}
\submitjournal{ApJL}

\shorttitle{Search for Astronomical Neutrinos from Blazar TXS0506+056 in Super-Kamiokande}
\shortauthors{K.Hagiwara et al.}

\begin{document}

\title{Search for Astronomical Neutrinos from Blazar TXS0506+056 in Super-Kamiokande}

\correspondingauthor{Kaito Hagiwara}
\email{k.hagiwara@s.okayama-u.ac.jp}

\newcommand{\AFFicrr}{\affiliation{Kamioka Observatory, Institute for Cosmic Ray Research, University of Tokyo, Kamioka, Gifu 506-1205, Japan}}
\newcommand{\AFFkashiwa}{\affiliation{Research Center for Cosmic Neutrinos, Institute for Cosmic Ray Research, University of Tokyo, Kashiwa, Chiba 277-8582, Japan}}
\newcommand{\AFFipmu}{\affiliation{Kavli Institute for the Physics and
Mathematics of the Universe (WPI), The University of Tokyo Institutes for Advanced Study,
University of Tokyo, Kashiwa, Chiba 277-8583, Japan }}
\newcommand{\AFFmad}{\affiliation{Department of Theoretical Physics, University Autonoma Madrid, 28049 Madrid, Spain}}
\newcommand{\AFFubc}{\affiliation{Department of Physics and Astronomy, University of British Columbia, Vancouver, BC, V6T1Z4, Canada}}
\newcommand{\AFFbu}{\affiliation{Department of Physics, Boston University, Boston, MA 02215, USA}}
\newcommand{\AFFuci}{\affiliation{Department of Physics and Astronomy, University of California, Irvine, Irvine, CA 92697-4575, USA }}
\newcommand{\AFFcsu}{\affiliation{Department of Physics, California State University, Dominguez Hills, Carson, CA 90747, USA}}
\newcommand{\AFFcnm}{\affiliation{Department of Physics, Chonnam National University, Kwangju 500-757, Korea}}
\newcommand{\AFFduke}{\affiliation{Department of Physics, Duke University, Durham NC 27708, USA}}
\newcommand{\AFFfukuoka}{\affiliation{Junior College, Fukuoka Institute of Technology, Fukuoka, Fukuoka 811-0295, Japan}}
\newcommand{\AFFgifu}{\affiliation{Department of Physics, Gifu University, Gifu, Gifu 501-1193, Japan}}
\newcommand{\AFFgist}{\affiliation{GIST College, Gwangju Institute of Science and Technology, Gwangju 500-712, Korea}}
\newcommand{\AFFuh}{\affiliation{Department of Physics and Astronomy, University of Hawaii, Honolulu, HI 96822, USA}}
\newcommand{\AFFicl}{\affiliation{Department of Physics, Imperial College London , London, SW7 2AZ, United Kingdom }}
\newcommand{\AFFkek}{\affiliation{High Energy Accelerator Research Organization (KEK), Tsukuba, Ibaraki 305-0801, Japan }}
\newcommand{\AFFkobe}{\affiliation{Department of Physics, Kobe University, Kobe, Hyogo 657-8501, Japan}}
\newcommand{\AFFkyoto}{\affiliation{Department of Physics, Kyoto University, Kyoto, Kyoto 606-8502, Japan}}
\newcommand{\AFFliv}{\affiliation{Department of Physics, University of Liverpool, Liverpool, L69 7ZE, United Kingdom}}
\newcommand{\AFFmiyagi}{\affiliation{Department of Physics, Miyagi University of Education, Sendai, Miyagi 980-0845, Japan}}
\newcommand{\AFFnagoya}{\affiliation{Institute for Space-Earth Environmental Research, Nagoya University, Nagoya, Aichi 464-8602, Japan}}
\newcommand{\AFFkmi}{\affiliation{Kobayashi-Maskawa Institute for the Origin of Particles and the Universe, Nagoya University, Nagoya, Aichi 464-8602, Japan}}
\newcommand{\AFFpol}{\affiliation{National Centre For Nuclear Research, 02-093 Warsaw, Poland}}
\newcommand{\AFFsuny}{\affiliation{Department of Physics and Astronomy, State University of New York at Stony Brook, NY 11794-3800, USA}}
\newcommand{\AFFokayama}{\affiliation{Department of Physics, Okayama University, Okayama, Okayama 700-8530, Japan }}
\newcommand{\AFFosaka}{\affiliation{Department of Physics, Osaka University, Toyonaka, Osaka 560-0043, Japan}}
\newcommand{\AFFox}{\affiliation{Department of Physics, Oxford University, Oxford, OX1 3PU, United Kingdom}}
\newcommand{\AFFqmul}{\affiliation{School of Physics and Astronomy, Queen Mary University of London, London, E1 4NS, United Kingdom}}
\newcommand{\AFFregina}{\affiliation{Department of Physics, University of Regina, 3737 Wascana Parkway, Regina, SK, S4SOA2, Canada}}
\newcommand{\AFFseoul}{\affiliation{Department of Physics, Seoul National University, Seoul 151-742, Korea}}
\newcommand{\AFFsheff}{\affiliation{Department of Physics and Astronomy, University of Sheffield, S3 7RH, Sheffield, United Kingdom}}
\newcommand{\AFFshizuokasc}{\affiliation{Department of Informatics in
Social Welfare, Shizuoka University of Welfare, Yaizu, Shizuoka, 425-8611, Japan}}
\newcommand{\AFFstfc}{\affiliation{STFC, Rutherford Appleton Laboratory, Harwell Oxford, and Daresbury Laboratory, Warrington, OX11 0QX, United Kingdom}}
\newcommand{\AFFskk}{\affiliation{Department of Physics, Sungkyunkwan University, Suwon 440-746, Korea}}
\newcommand{\AFFtokyo}{\affiliation{The University of Tokyo, Bunkyo, Tokyo 113-0033, Japan }}
\newcommand{\AFFtodai}{\affiliation{Department of Physics, University of Tokyo, Bunkyo, Tokyo 113-0033, Japan }}
\newcommand{\AFFtit}{\affiliation{Department of Physics,Tokyo Institute of Technology, Meguro, Tokyo 152-8551, Japan }}
\newcommand{\AFFtus}{\affiliation{Department of Physics, Faculty of Science and Technology, Tokyo University of Science, Noda, Chiba 278-8510, Japan }}
\newcommand{\AFFtoronto}{\affiliation{Department of Physics, University of Toronto, ON, M5S 1A7, Canada }}
\newcommand{\AFFtriumf}{\affiliation{TRIUMF, 4004 Wesbrook Mall, Vancouver, BC, V6T2A3, Canada }}
\newcommand{\AFFtokai}{\affiliation{Department of Physics, Tokai University, Hiratsuka, Kanagawa 259-1292, Japan}}
\newcommand{\AFFtsinghua}{\affiliation{Department of Engineering Physics, Tsinghua University, Beijing, 100084, China}}
\newcommand{\AFFynu}{\affiliation{Faculty of Engineering, Yokohama National University, Yokohama, Kanagawa, 240-8501, Japan}}
\newcommand{\AFFllr}{\affiliation{Ecole Polytechnique, IN2P3-CNRS, Laboratoire Leprince-Ringuet, F-91120 Palaiseau, France }}
\newcommand{\AFFbari}{\affiliation{ Dipartimento Interuniversitario di Fisica, INFN Sezione di Bari and Universit\`a e Politecnico di Bari, I-70125, Bari, Italy}}
\newcommand{\AFFnapoli}{\affiliation{Dipartimento di Fisica, INFN Sezione di Napoli and Universit\`a di Napoli, I-80126, Napoli, Italy}}
\newcommand{\AFFroma}{\affiliation{INFN Sezione di Roma and Universit\`a di Roma ``La Sapienza'', I-00185, Roma, Italy}}
\newcommand{\AFFpadova}{\affiliation{Dipartimento di Fisica, INFN Sezione di Padova and Universit\`a di Padova, I-35131, Padova, Italy}}
\newcommand{\AFFkeio}{\affiliation{Department of Physics, Keio University, Yokohama, Kanagawa, 223-8522, Japan}}
\newcommand{\AFFwinnipeg}{\affiliation{Department of Physics, University of Winnipeg, MB R3J 3L8, Canada }}
\newcommand{\AFFkcl}{\affiliation{Department of Physics, King's College London, London,WC2R 2LS, UK }}

\AFFokayama

\AFFicrr
\AFFkashiwa
\AFFmad
\AFFbu
\AFFuci
\AFFcsu
\AFFcnm
\AFFduke
\AFFllr
\AFFfukuoka
\AFFgifu
\AFFgist
\AFFuh
\AFFicl
\AFFbari
\AFFnapoli
\AFFpadova
\AFFroma
\AFFkcl
\AFFkeio
\AFFkek
\AFFkobe
\AFFkyoto
\AFFliv
\AFFmiyagi
\AFFnagoya
\AFFkmi
\AFFpol
\AFFsuny

\AFFosaka
\AFFox
\AFFseoul
\AFFsheff
\AFFshizuokasc
\AFFstfc
\AFFskk
\AFFtokai
\AFFtokyo
\AFFtodai
\AFFipmu
\AFFtit
\AFFtus
\AFFtoronto
\AFFtriumf
\AFFtsinghua
\AFFwinnipeg
\AFFynu

\author{K.~Hagiwara}
\AFFokayama
\author{K.~Abe}
\AFFicrr
\AFFipmu
\author{C.~Bronner}
\AFFicrr
\author{Y.~Hayato}
\AFFicrr
\AFFipmu
\author{M.~Ikeda}
\AFFicrr
\author{H.~Ito}
\AFFicrr 
\author{J.~Kameda}
\AFFicrr
\AFFipmu
\author{Y.~Kataoka}
\AFFicrr
\author{Y.~Kato}
\AFFicrr
\author{Y.~Kishimoto}
\AFFicrr
\AFFipmu 
% ...


\collaboration{The Super-Kamiokande Collaboration}
\noaffiliation



\begin{abstract}


We report a search for astronomical neutrinos in the energy region from several GeV to TeV in the direction of the blazar TXS0506+056
using the Super-Kamiokande detector following the detection of a 100~TeV neutrino from the same location 
by the IceCube collaboration.
Using Super-Kamiokande neutrino data across several data samples observed from April 1996 to February 2018 
we have searched for both a total excess above known backgrounds across the entire period as well as 
localized excesses on smaller time scales in that interval.
No significant excess nor significant variation in the observed event rate are found in the blazar direction.
Upper limits are placed on the electron and muon neutrino fluxes at 90\% confidence level as $6.0 \times 10^{-7}$ and $4.5 \times 10^{-7}$ to $9.3 \times 10^{-10}$ [${\rm erg}/{\rm cm}^2/{\rm s}$], respectively.

\end{abstract}



\section{Introduction}

TXS0506+056 is a BL Lac type blazar (redshift $z=0.3365 \pm 0.0010$~\citep{Paiano_2018}) 
and located at right ascension (R.A.)=$77.3582^\circ$ and declination (Dec.)=$+5.6931^\circ$ (J2000 equinox)~\citep{Massaro_2015}. 
The IceCube Neutrino Observatory~\citep{Aartsen_2017} detected a high-energy neutrino event with an estimated energy of 290 TeV on 22 September, 2017 at 20:54:30.43 Coordinated Universal Time (IceCube-170922A), the arrival direction of which coincides 
with the location of TXS0506+056~\citep{eaat1378, icecube18}.
Within a minute of detection, this event's information shared via the Gamma-ray Coordinate Network (GCN) (GCN/AMON Notice: \url{https://gcn.gsfc.nasa.gov/amon.html}) and follow-up observations over a wide range of energiers were carried out by several observatories.
According to the Fermi All-Sky Variability Analysis (FAVA)~\citep{Abdollahi_2017}, 
TXS0506+056 brightened in the GeV band starting in April 2017~\citep{2017ATel10791}.
Fermi's Automated Science Processing (ASP) also found a gamma-ray flare from this source years before. 
Subsequently, the IceCube collaboration additionally reported a possible neutrino event excess 
from this blazar in older data between September 2014 and March 2015~\citep{icecube18}.
Coincidence between the neutrino arrival direction and the blazar location as well as timing correlated with the associated gamma-ray flare 
suggest that the observed neutrinos originated from the blazar and strongly motivate searches for neutrinos in the other energy regions.

\section{Super-Kamiokande experiment} 

Super-Kamiokande (SK) \citep{FUKUDA2003418} is a large water Cherenkov detector located 1000~m underground (2700 m.w.e.) 
in the Kamioka-mine, Gifu Prefecture, Japan. 
It is a cylindrical detector, 39.3~m in diameter and 41.4~m in height and contains 
50~kilotonnes of ultra-pure water as neutrino target.
The tank is optically separated into an inner detector (ID) and an outer detector (OD) by a structure placed $\sim$2~m 
from the tank wall.
More than 11,000 20-inch photomultiplier tubes (PMTs) in the ID are used to observe 
the pattern and amount of Cherenkov photons emitted by charged particles produced by neutrino interactions in the water.
The OD is primarily used as a veto and has 1885 8-inch PMTs and is covered with a Tyvek sheet to enhance the light reflection from Cherenkov photons.
Super-Kamiokande is primarily sensitive to particle interactions in the energy region of several MeV to a few tens of TeV.
In particular, SK observes atmospheric neutrinos above several 10~MeV at a rate of $\sim$8 events per day 
in a 22.5~kilotonne fiducial volume within the ID~\citep{Jiang_2019}.
The overburden of the mountain above the detector reduces the cosmic ray muon rate at the detector 
by a factor of $\sim$$10^{5}$ compared to that at the surface.
Such backgrounds are almost completely eliminated by anti-coincidence of the ID and OD, reducing non-neutrino events to less than 1\% of the final data sample.

The SK experiment has been operated since April 1996 and has made observations in four distinct phases known as SK-I, SK-II, SK-III, and SK-IV.
The first phase, SK-I, lasted from April 1996 to July 2001 (1489.2 livetime days) with 40\% photocoverage of the ID using 11146 PMTs.
In November 2001, half of those PMTs were lost to an accident following detector maintenance, so the SK-II phase was 
operated from December 2002 to October 2005 (798.6 livetime days) with a reduced photocoverage of 19\% (5182 PMTs).
After replacing the missing PMTs in April 2006, the SK-III period operated with the full photocoverage (11129 PMTs) until September 2008 (518.1 livetime days).
New front end electronics were installed immediately thereafter to start the SK-IV phase. 
This period lasted until June 2018, when refurbishment work ahead of a detector upgrade began.
Though the detector is currently running as SK-V, in this paper neutrino data from SK-I to SK-IV through February 2018 corresponding to  5924.4 live days are used for analysis.

\section{Event sample}

The present analysis utilizes the Super-Kamiokande neutrino data with more than 100~MeV of visible energy, divided
into three classes depending upon the event topology.
In the fully-contained (FC) and partially-contained (PC) event samples, the neutrino interaction is reconstructed within the ID using Cherenkov rings produced by its daughter particles. 
An event where all daughter particles stop inside the ID is classified as FC and those where at least one particle exits the ID
and deposits energy in the OD is classified as PC.
Upward-going muon (UPMU) events are observed when energetic muons produced by muon-neutrino interactions with the rock surrounding the detector penetrate the ID from below its horizon.
Since similar downward-going neutrino events suffer from a large amount of cosmic ray muon backgrounds 
from above the detector, the event direction of the UPMU sample is restricted to be upward. 
There is no such restriction on FC and PC events.
Both electron neutrinos ($\nu_e$, $\bar{\nu}_e$) and muon neutrinos ($\nu_\mu$, $\bar{\nu}_\mu$) are observed in the FC sample while only muon neutrinos populate the PC and UPMU samples.

Events with vertices inside the fiducial volume, defined as the region in the ID more than 2.0~m from any wall, are selected for the FC and PC samples.
Separation between the FC and PC samples is determined by the number of effective PMT hits in OD; 
the FC sample requires fewer than 16 hits (10 hits for SK-I). 
Events rejected by this cut are classified as PC. 
UPMU events where the muon passes through the detector (upward through-going muons) as well as events where it stops in the detector (upward stopping muons) are included in this analysis.
Upward through-going muon events are required to have a muon track longer than 7.0~m 
and stopping events are required to have a muon momentum greater than 1.6~GeV.
Both criteria ensure that the reconstructed muon direction is from below the horizon.
Details of the event selection can be found in \citet{Ashie_2005}.

In order to estimate the atmospheric neutrino background for this search, a 500-year-equivalent Monte Carlo (MC) simulation of each SK phase has been used. 
The NEUT interaction generator~\citep{2009AcPPB..40.2477H} has been adopted for interactions in water 
and a detector simulation based on the Geant3~\citep{Brun:1119728} framework is employed for tracking secondary particles and simulating the detector response.
Additional simulation details are presented in ~\citet{Abe_2018}

The arrival direction of neutrinos is determined by reconstruction of Cherenkov rings in the ID and the reconstruction quality typically depends on 
the number of such rings and the energy of the initial neutrino interaction.
In order to use events with sufficient angular resolution for association with the blazar direction in this search,
an additional cut on the observed energy is applied, 5.1~GeV for FC events and 1.8~GeV for PC events.
This ensures that the angular deviation of the reconstructed direction from the truth is within 10~degrees for more than 68\% of these events.
Since UPMU events originate from neutrinos with higher energy than other categories, 
their arrival direction is estimated with higher accuracy. 
Therefore no additional restriction on the UPMU energy is used as more than 77\% of events are reconstructed within 5~degrees 
of the true arrival direction.

\section{Analysis results}

The analysis described below searches for a possible neutrino excess from the blazar by first counting the number of neutrinos in an angular region around the direction to the assumed source.
Then the number of events coming from an alternative direction are studied to check the consistency of the observation and background predictions.
Finally, a simple statistical method is used to test for local increases in the event rate coming from the blazar direction.

\begin{figure}[ht!]
	\plotone{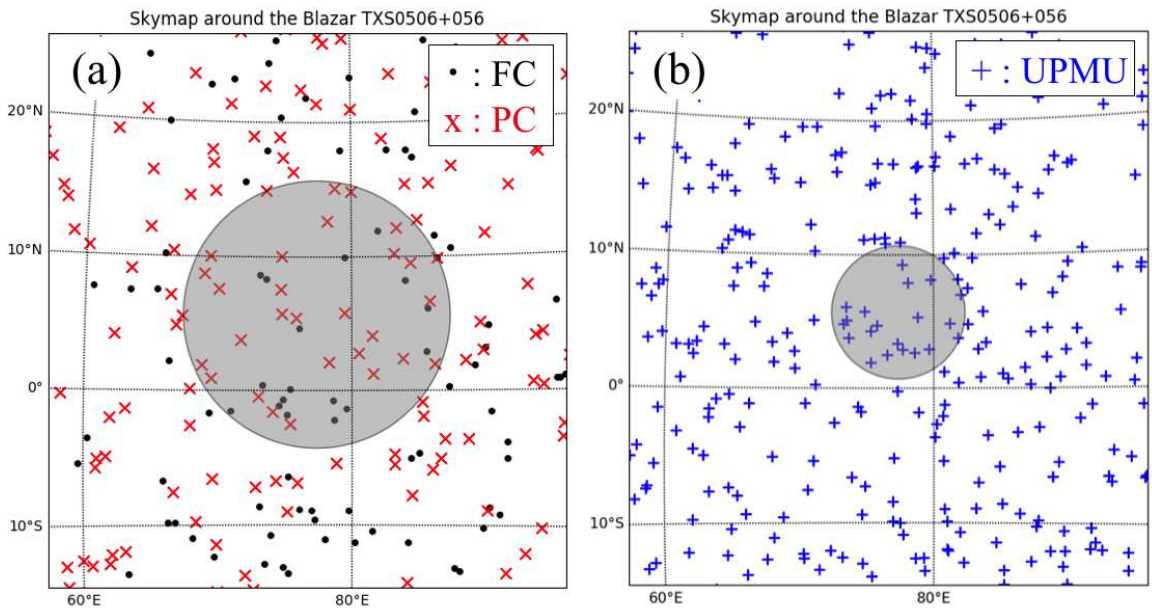}
	\caption{
	Reconstructed arrival directions of fully-contained (FC, black circle), partially-contained (PC, red x), and upward-going muons (UPMU, blue +) events around the location of blazar TXS0506+056 $(\alpha, \beta)=(77.3582^\circ, +5.6931^\circ)$ in equatorial coordinates.  The horizontal axis is the right ascension and the veritical axis the declination.
	The shaded circle in the left (right) figure shows the 10 (5) degree search cone used in the analysis of FC and PC (UPMU) events.
	\label{fig1}}
\end{figure}

Figure~\ref{fig1} shows a sky map of the reconstructed arrival direction of selected neutrino events around the blazar direction for all samples.
Based on the angular resolution of the reconstructed direction in each sample, we defined the search region to be 10 (5) degrees around the blazar location for FC and PC (UPMU)~\citep{Abe_2017}.
There were 18 FC, 29 PC, and 20 UPMU events observed in these regions during SK~I--IV.
Note that neutrino events observed in the FC sample include both electron neutrino ($\nu_e$ and $\bar{\nu}_e$) and muon neutrino ($\nu_\mu$ and $\bar{\nu}_\mu$) interactions.

In order to quantitatively study possible event excesses above atmospheric neutrino backgrounds, 
MC is used to predict the event rate in the search region of each sample. 
The atmospheric neutrino event rate depends on the arrival direction because the thickness of the atmosphere 
and the neutrino oscillation probability change with zenith angle (in detector coordinates). 
The thicker the atmosphere, the higher the probability that atmospheric neutrinos will be generated.
For downward- or horizontally-produced neutrinos the path length to the detector is relatively short 
and the effect of neutrino oscillations is reduced.
Consequently, since the zenith angle is related to declination, the event rate also varies with declination.
To simulate the effect of varying right ascension in the actual data, MC events are randomly assigned 
right ascension values under the assumption of a flat local sidereal time.
Several corrections are applied on an event-by-event basis to account for neutrino oscillations and to 
reflect best-fit values of systematic error parameters from the analysis in~\citet{Abe_2018}.
The event rates in the search regions are then calculated for each SK phase and combined with appropriate livetime normalization factors.

\begin{figure}[ht!]
\plotone{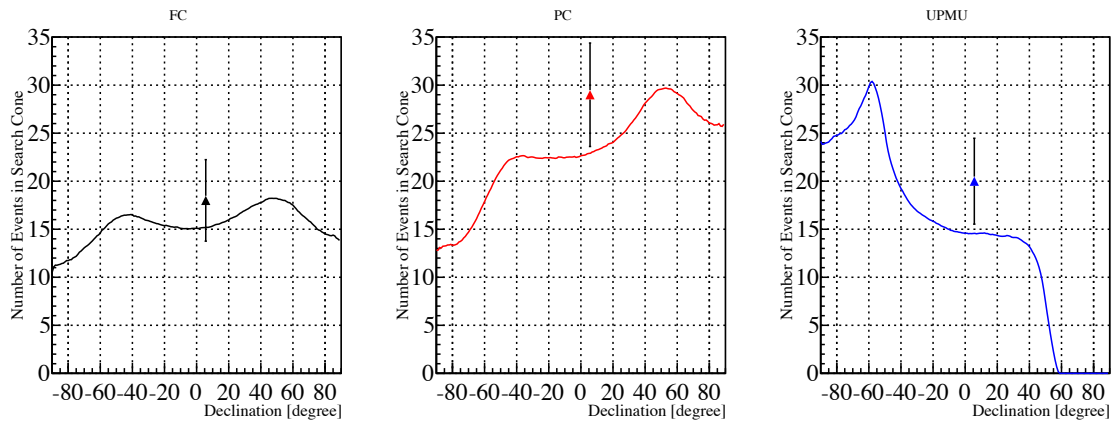}
\caption{Number of detected events in each search region (points with error bar) are shown with corresponding predictions for the FC (left, black), PC (middle, red), and UPMU (right, blue) samples.
	The error bar shows the statistical error.
	\label{fig2}}
\end{figure}

Figure~\ref{fig2} shows the observed data events in the each fixed search cone superimposed on the MC taken over various declinations.
Note that the FC and PC background events distribute across all declinations and show a slight increase at higher declinations, especially PC events, due to the decrease of upward-going muon neutrinos lost to neutrino oscillations.
The double-bump structure around $-50$ and $50$ degrees is due to the increased atmospheric neutrino flux coming 
from the near$-$horizon direction, where the effective atmospheric depth is deeper than other directions.
Since UPMU events are required to come from below the horizon, their maximum declination is about 54 degrees.
The observed data agrees with the expected background within 0.7$\sigma$ for FC, 1.1$\sigma$ for PC, and 1.2$\sigma$ for UPMU events, considering statistical uncertainties alone.

To further check the consistency of the observed event rates in the search cones, we further investigate by studying similarly sized angular regions taken at the same declination as TXS0506+056 but with different right ascension values.  
The average and variance of the number of observed events in these ``off-source'' regions are compared with thse in the ``on-source'' region around the blazar.
They are consistent within 0.5$\sigma$ for FC, 1.6$\sigma$ for PC, and 1.5$\sigma$ for UPMU based on counting statistics only.
The event rate in the on-source search cone is $3.0\pm0.7$ for FC, $4.9\pm0.9$ for PC, and $3.4\pm0.8$ for UPMU events per 1000 livetime days.
Averaging the off-source rates yields $2.7\pm0.6$ for FC, $3.9\pm0.6$ for PC, and $2.5\pm0.6$ events per 1000 livetime days.
Therefore the on-source and off-source rates are consistent, indicating no excess of neutrino events in the direction of the blazar.

\begin{figure}[ht!]
	\plotone{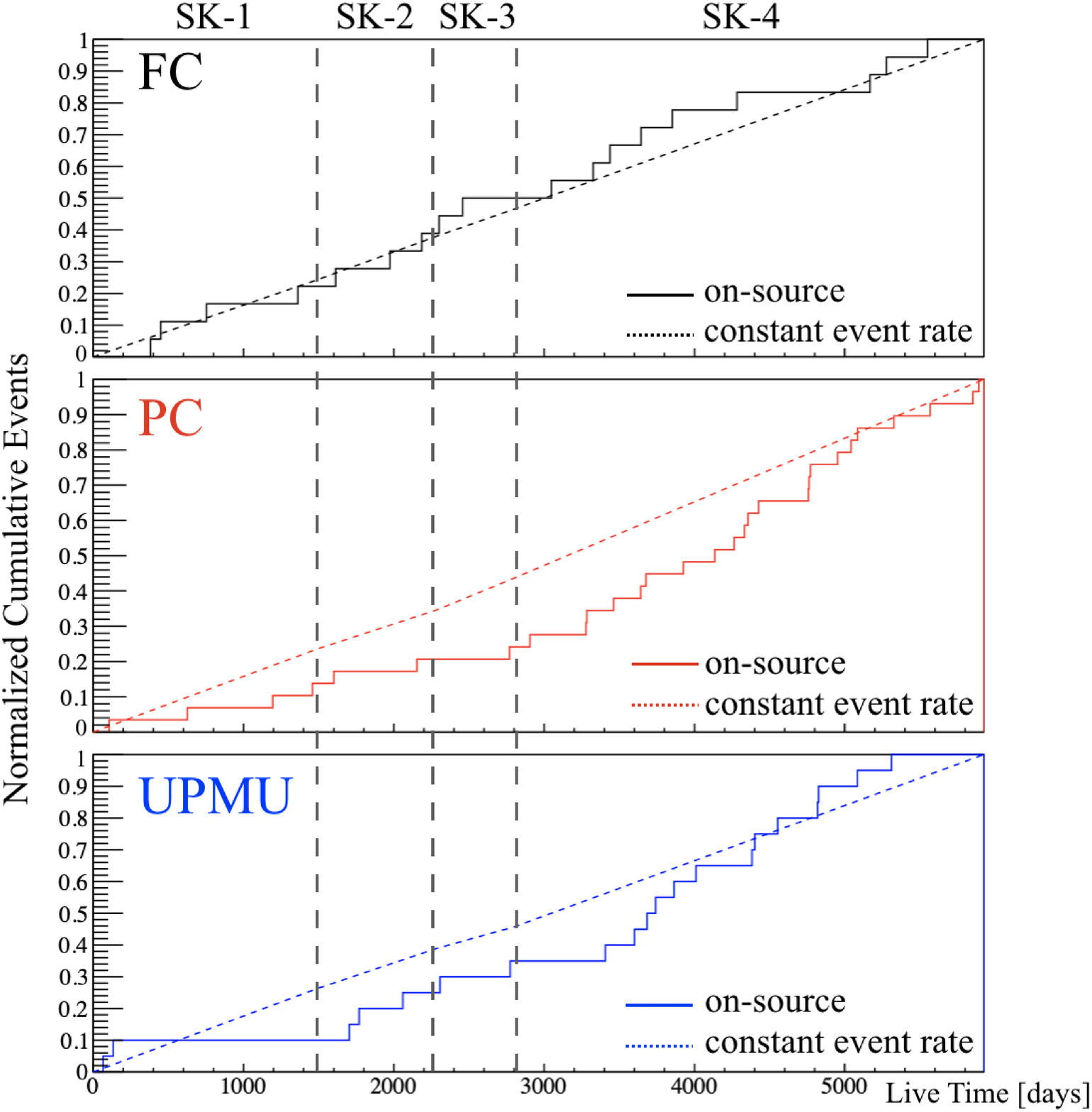}
	\caption{Normalized cumulative events as a function of livetime day for FC, PC, and UPMU.
	The solid lines are observed events from the on-source region and the dashed lines are estimated background events assuming a constant event rate for each SK phase.
	The ranges of each SK phase are shown as vertical dashed lines.
	The maximum distance between the experimental data and the expectation is 0.13 for FC, 0.25 for PC, and 0.21 for UPMU.
	\label{fig3}}
\end{figure}

We additionally searched for evidence of a local increase in the neutrino event rate in the period April 1996 to February 2018 to test for possible correlation with gamma-ray flaring of the blazar.
Since the atmospheric neutrino rate is known to be stable at each SK phase, the number of observed neutrino events is expected to increase linearly with increasing livetime if there is no neutrino emission from the blazar. 
On the other hand, the event rate would deviate from linearity if there were additional neutrinos from the gamma-ray flare.
In order to test for the presence of such a variation, we evaluated the probability ($p$-value) that the observed rate is consistent using a Kolmogorov-Smirnov test (KS-test).
Figure~\ref{fig3} compares the cumulative observed events with the expected events as a function of livetime day.
To estimate the degree of deviation, a set of 10,000 pseudo experiments was generated assuming that the expected background from the MC in each SK phase was distributed according to a Poisson function.
In each pseudo experiment, the observed time of each event was randomly assigned assuming a constant rate in each SK phase.
The maximum distance between each pseudo experiment and the expectation is compared to that from the observed SK data to calculate a $p$-value.
This represents the percentage of pseudo experiments that have a maximum distance larger than the data.
The estimated $p$-values are 91.74\%, 12.26\%, and 48.75\% for FC, PC, and UPMU events, respectively, indicating consistency with a constant event rate.
Accordingly, we conclude that no significant signal from the direction of blazar TXS0506+056 exists 
in the SK data during the observation period considered here.

Since no significant indication of a signal from the blazar was found in any of the tests above, we estimate flux limits based on the expected background throughout the entire observation period.
In the following we derive limits on the neutrino fluence from this blazar.
Based on the assumption of no signal, the upper limit on the neutrino fluence is estimated following~\citet{Swanson_2006} and \citet{Thrane_2009}:
\begin{eqnarray}
\Phi_{\rm FC}^{\nu_x+\bar{\nu}_x} &=& \frac{N_{90}^{\rm FC}}{N_T \int {\rm d}E_\nu \left( \sigma^{\nu_x} (E_\nu) \varepsilon^{\nu_x} (E_\nu) + \sigma^{\bar{\nu}_x} (E_\nu) \varepsilon^{\bar{\nu}_x} (E_\nu) \right) \lambda(E_\nu^{-2})}~~~~~(x=e,\mu), \\
\Phi_{\rm PC}^{\nu_\mu+\bar{\nu}_\mu} &=& \frac{N_{90}^{\rm PC}}{N_T \int {\rm d}E_\nu \left( \sigma^{\nu_\mu}(E_\nu) \varepsilon^{\nu_\mu}(E_\nu) + \sigma^{\bar{\nu}_\mu}(E_\nu) \varepsilon^{\bar{\nu}_\mu}(E_\nu) \right) \lambda(E_\nu^{-2})}, \\
\Phi_{\rm UPMU}^{\nu_\mu+\bar{\nu}_\mu} &=& \frac{N_{90}^{\rm UPMU}}{A_{\rm eff}(z) \int {\rm d}E_\nu \left( P^{\nu_\mu}(E_\nu) S^{\nu_\mu}(z, E_\nu) + P^{\bar{\nu}_\mu}(E_\nu) S^{\bar{\nu}_\mu}(z, E_\nu) \right) \lambda(E_\nu^{-2})}.
\label{eq:fluence}
\end{eqnarray}
\noindent
Here $N_{90}$ is the upper limit on the number of the events above the background at 90\% Confidence Level (C.L.), 
$N_T$ is the number of nucleons in the 22.5 kiloton fiducial volume of the detector,
$\sigma(E_\nu)$ is the total neutrino interaction cross-section from the NEUT model, 
and $\varepsilon(E_\nu)$ is the neutrino detection efficiency.
The parameter $A_{\rm eff}$ is the effective area of SK to UPMU interactions 
and $P(E_\nu)$ is the probability for a neutrino to produce a muon in the rock surrounding that reaches SK. 
Calculation of the latter is done using the charged-current cross section for muon neutrino-nucleon interaction in rock coupled with the expected range of the resulting muon produced assuming the initial vertex is some distance from the detector. 
The attenuation of neutrinos due to their interactions in the earth is given by $S(z, E_\nu)$ and taken from \citet{Gandhi_1996} using the ``Preliminary Earth Model"~\citep{DZIEWONSKI1981297}.
Finally, $\lambda (E_\nu^{-2})$ is the number density distribution from the blazar's direction, which we assume to follow a power law with spectral index of $-2$ as in the IceCube analysis~\citep{eaat1378}.

Assuming Poisson statistics for the expected backgrounds described above, $N_{90}$ for the FC, PC, and UPMU samples has been estimated to be 10.2, 14.6, and 12.7, respectively.
For the FC sample, we conservatively make no distinction between electron and muon type neutrinos when calculating $N_{90}$.\footnote{Though tau neutrinos are present in the SK data, they represent a negligible contribution to the current data set.}
The energy ranges used in the integrals in Equation~\ref{eq:fluence} are 5.1 - 10~GeV (FC), 1.8 - 100~GeV (PC), and 1.6~GeV - 10~TeV (UPMU). 
These ranges represent the MC neutrino energies populating each sample.
Upper limits are calculated for both electron and muon neutrinos using FC events since this sample is sensitive to both.
Only that of muon neutrino fluence limits are estimated for the other samples.
The results are shown in Table~\ref{tab:fluence}.
In these calculations, the fluence limits are calculated using the average zenith angle
to the source taken over the detector observation period.
since the effective area, $A_{\rm eff}(z)$, and the shadowing effect, $S(z, E_\nu)$, depend on the zenith angle and fluctuate with the motion of the Earth.

\begin{deluxetable}{cccc}[h!]
\tablecaption{
Summary of the data samples and fluence limits.
Upper limits at 90\% C.L. on the neutrino fluence have been  
calculated assuming an $E^{-2}$ energy spectrum from blazar TXS0506+056.
\label{tab:fluence}
}
\tablecolumns{8}
\tablewidth{0pt}
\tablehead{
\colhead{} & 
\colhead{FC} &
\colhead{PC} &
\colhead{UPMU} 
}
\startdata
Energy Range [GeV] & 5.1-10 & 1.8-100 & 1.6-$10^4$ \\
Search Cone [$^\circ$] & 10 & 10 & 5 \\
Observed Events & 18 & 29 & 20 \\
Expected Background & 15.2 & 22.9 & 14.5 \\
$N_{90}$ & 10.2 & 14.6 & 12.7 \\
Fluence Limit ($\nu_{\mu} + \bar{\nu}_{\mu}$) [${\rm cm}^{-2}$] & $6.9\times10^4$ & $1.1\times10^5$ & $1.3\times10^2$ \\
Fluence Limit ($\nu_{e} + \bar{\nu}_{e}$) [${\rm cm}^{-2}$] & $1.9\times10^4$ & - & - 
\enddata
\end{deluxetable}

\begin{figure}[ht!]
\plotone{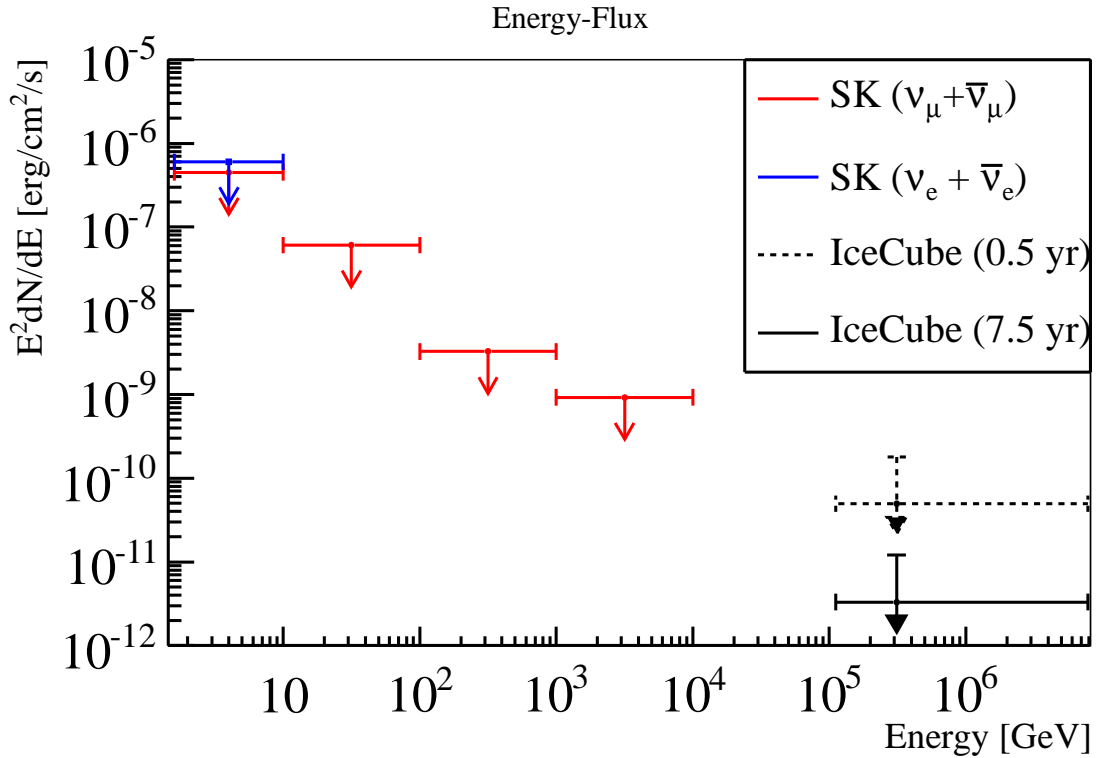}
\caption{
90\% C.L. energy-dependent flux upper limit in the direction of blazar TXS0506+056 by SK $\nu_\mu+\bar{\nu}_\mu$ (red) and $\nu_e+\bar{\nu}_e$ (blue) compared with IceCube~\citep{eaat1378}.
\label{fig4}
}
\end{figure}

Figure~\ref{fig4} shows the energy-dependent upper limits for electron neutrinos ($\nu_e+\bar{\nu}_e$) and muon neutrinos ($\nu_\mu+\bar{\nu}_\mu$) fluxes by SK observations compared IceCube group~\citep{eaat1378}. 
The IceCube collaboration considered two neutrino emission periods to calculate the flux limit. 
In the first scenario, neutrinos were assumed to be emitted only during the about 6 month period corresponding to the duration of the $\gamma$-ray flare.
Alternatively, neutrinos emitted over the whole observation of IceCube (7.5 years) were considered.
These two benchmark cases and the result of our analysis are shown in Figure~\ref{fig4}.

Above 10~GeV the upper limits are obtained from UPMU data by using same formula described above with the corresponding energy range of the integration.
It should be noted that the neutrino energy cannot be reconstructed for UPMU events since they are produced by neutrinos interacting in the rock surrounding the detector.
Therefore, the same value of $N_{90}$ is used to calculate the upper limit in each energy bin.
Note that the FC sample is populated almost entirely by events with energy less than 10 GeV, so our limit for 
electron neutrinos spans a single bin.
For muon neutrinos in the 1 to 10~GeV bin and in the 10 to 100~GeV bin, 
the limit is calculated using the total observation and expectation, summing over the FC, PC, and UPMU contributions. 
%
%
The flux limit by SK covers $\sim$1~GeV to 10~TeV, and becomes more stringent as the energy increase because of larger neutrino cross section and smaller atmospheric neutrino backgrounds.

We note that the IceCube group observed evidence of a neutrino event excess between 2014 and 2015 from the direction of the blazar 
whose best fit energy spectrum was $E^{-2.2\pm 0.2}$ and whose flux was $2.5^{+1.1}_{-1.0}\times 10^{-9}\ [{\rm erg/cm^2/s}]$ at 100~TeV.
This corresponds to a flux of 
$2.0 \times 10^{-8} [{\rm erg/cm^2/s}]$ at 3.2~GeV 
and 
$5.1 \times 10^{-9} [{\rm erg/cm^2/s}]$ at 3.2~TeV, respectively. 
They can be compared limits from the present analysis of 
$4.5 \times 10^{-7} [{\rm erg/cm^2/s}]$ and 
$9.3 \times 10^{-10} [{\rm erg/cm^2/s}]$ 
in the respective energy regions.
If the IceCube spectral index is used in our calculations, the flux limits increase by 15\% at most.

\section{Conclusion}

We performed a search for neutrino detections coincident in the direction of blazar TXS0506+056 using the observation data from April 1996 to February 2018 by Super-Kamiokande in GeV to several TeV regions.
By comparing to the expected backgrounds, no significance excess was observed at greater than the 2$\sigma$ level in the blazar direction of 5 (10) degrees for UPMU (FC, PC) samples.
No significant temporal increase of neutrino flux was found in the blazar direction by examining the change of the event rate using the Kolmogorov-Smirnov test.
Based on no signal assumption, upper limits of neutrino fluence and the energy-dependent neutrino flux are given for both electron neutrinos and muon neutrinos.
Upper limits are placed on the electron neutrino flux of $6.0 \times 10^{-7} [{\rm erg}/{\rm cm}^2/{\rm s}$ below 10 GeV 
and on the muon neutrino flux of 
$4.5 \times 10^{-7} [{\rm erg}/{\rm cm}^2/{\rm s}]$ to 
$9.3 \times 10^{-10}$ [${\rm erg}/{\rm cm}^2/{\rm s}$] in the range 1~GeV to 10~TeV 
assuming an $E^{-2}$ energy spectrum.


\acknowledgments

We gratefully acknowledge the cooperation of the Kamioka Mining and Smelting Company.
The Super‐Kamiokande experiment has been built and operated from funding by the 
Japanese Ministry of Education, Culture, Sports, Science and Technology, the U.S.
Department of Energy, and the U.S. National Science Foundation. Some of us have been 
supported by funds from the National Research Foundation of Korea NRF‐2009‐0083526
(KNRC) funded by the Ministry of Science, ICT, and Future Planning and the the Ministry of
Education (2018R1D1A3B07050696, 2018R1D1A1B07049158), 
the Japan Society for the Promotion of Science, the National
Natural Science Foundation of China under Grants No. 11235006, the National Science and 
Engineering Research Council (NSERC) of Canada, the Scinet and Westgrid consortia of
Compute Canada, the National Science Centre, Poland (2015/18/E/ST2/00758),
the Science and Technology Facilities Council (STFC) and GridPPP, UK, 
and the European Union’s H2020-MSCA-RISE-2018 JENNIFER2 grant agreement no.822070.

\bibliographystyle{aasjournal}
\bibliography{blazar}{}






\end{document}